\documentclass[dvipsnames]{INTERSPEECH2023}
\usepackage{tabto}
\usepackage[nolist]{acronym}%
\usepackage{float, stfloats}
\usepackage{subfig}
\usepackage{xcolor}
\usepackage{graphicx}

\usepackage{tikz}
\usetikzlibrary{arrows.meta,patterns}
\usetikzlibrary{ipe} 
\usepackage[vskip=0em,leftmargin=1em,rightmargin=1em]{quoting}

\usepackage{scalefnt}
\usepackage{makecell}
\usepackage{booktabs}
\usepackage{multirow}
\usepackage{fontawesome5}
\usepackage{todonotes}
\usepackage[style=ieee, doi=false, isbn=false, url=false, eprint=false, related=false, maxnames=5, minnames=1, date=year]{biblatex}
\usepackage{xurl}
\AtEveryBibitem{\clearlist{language}}
\AtEveryBibitem{\clearfield{note}}
\AtEveryBibitem{\clearfield{address}}
\AtEveryBibitem{\clearlist{address}}
\AtEveryBibitem{\clearlist{location}}
\AtEveryBibitem{\clearfield{location}}
\AtEveryBibitem{\clearfield{publisher}}
\AtEveryBibitem{\clearlist{publisher}}
\AtEveryBibitem{\clearlist{pages}}
\AtEveryBibitem{\clearfield{pages}
    \clearfield{urlyear}
    \clearfield{urlmonth}}
\graphicspath{{./figures}}

\usepackage{comment}
\includecomment{subjeval}
\includecomment{torchsquim}

\newcommand{\ua}{$\uparrow$}
\newcommand{\da}{$\downarrow$}

\usepackage{scrextend}
\tikzstyle{ipe stylesheet} = [
  ipe import,
  even odd rule,
  line join=round,
  line cap=butt,
  ipe pen normal/.style={line width=0.4},
  ipe pen heavier/.style={line width=0.8},
  ipe pen fat/.style={line width=1.2},
  ipe pen ultrafat/.style={line width=2},
  ipe pen normal,
  ipe mark normal/.style={ipe mark scale=3},
  ipe mark large/.style={ipe mark scale=5},
  ipe mark small/.style={ipe mark scale=2},
  ipe mark tiny/.style={ipe mark scale=1.1},
  ipe mark normal,
  /pgf/arrow keys/.cd,
  ipe arrow normal/.style={scale=7},
  ipe arrow large/.style={scale=10},
  ipe arrow small/.style={scale=5},
  ipe arrow tiny/.style={scale=3},
  ipe arrow normal,
  /tikz/.cd,
  ipe arrows, 
  <->/.tip = ipe normal,
  ipe dash normal/.style={dash pattern=},
  ipe dash dotted/.style={dash pattern=on 1bp off 3bp},
  ipe dash dashed/.style={dash pattern=on 4bp off 4bp},
  ipe dash dash dotted/.style={dash pattern=on 4bp off 2bp on 1bp off 2bp},
  ipe dash dash dot dotted/.style={dash pattern=on 4bp off 2bp on 1bp off 2bp on 1bp off 2bp},
  ipe dash normal,
  ipe node/.append style={font=\normalsize},
  ipe stretch normal/.style={ipe node stretch=1},
  ipe stretch normal,
  ipe opacity 10/.style={opacity=0.1},
  ipe opacity 30/.style={opacity=0.3},
  ipe opacity 50/.style={opacity=0.5},
  ipe opacity 75/.style={opacity=0.75},
  ipe opacity opaque/.style={opacity=1},
  ipe opacity opaque,
]
\definecolor{red}{rgb}{1,0,0}
\definecolor{blue}{rgb}{0,0,1}
\definecolor{green}{rgb}{0,1,0}
\definecolor{yellow}{rgb}{1,1,0}
\definecolor{orange}{rgb}{1,0.647,0}
\definecolor{gold}{rgb}{1,0.843,0}
\definecolor{purple}{rgb}{0.627,0.125,0.941}
\definecolor{gray}{rgb}{0.745,0.745,0.745}
\definecolor{brown}{rgb}{0.647,0.165,0.165}
\definecolor{navy}{rgb}{0,0,0.502}
\definecolor{pink}{rgb}{1,0.753,0.796}
\definecolor{seagreen}{rgb}{0.18,0.545,0.341}
\definecolor{turquoise}{rgb}{0.251,0.878,0.816}
\definecolor{violet}{rgb}{0.933,0.51,0.933}
\definecolor{darkblue}{rgb}{0,0,0.545}
\definecolor{darkcyan}{rgb}{0,0.545,0.545}
\definecolor{darkgray}{rgb}{0.663,0.663,0.663}
\definecolor{darkgreen}{rgb}{0,0.392,0}
\definecolor{darkmagenta}{rgb}{0.545,0,0.545}
\definecolor{darkorange}{rgb}{1,0.549,0}
\definecolor{darkred}{rgb}{0.545,0,0}
\definecolor{lightblue}{rgb}{0.678,0.847,0.902}
\definecolor{lightcyan}{rgb}{0.878,1,1}
\definecolor{lightgray}{rgb}{0.827,0.827,0.827}
\definecolor{lightgreen}{rgb}{0.565,0.933,0.565}
\definecolor{lightyellow}{rgb}{1,1,0.878}
\definecolor{black}{rgb}{0,0,0}
\definecolor{white}{rgb}{1,1,1}

\interspeechcameraready 

\title{Evaluation of the Speech Resynthesis Capabilities \\of the VoicePrivacy Challenge Baseline B1}
\name{Ünal Ege Gaznepoglu, Nils Peters}
\address{
  International Audio Laboratories\thanks{\hspace{-1.8em}The International Audio Laboratories Erlangen are a joint institution of the University of Erlangen-Nürnberg and Fraunhofer IIS.}, Friedrich-Alexander-Universität Erlangen-Nürnberg, Germany}
\email{\{ege.gaznepoglu, nils.peters\} @fau.de}

\begin{document}

\maketitle
\newacro{BN}{bottleneck feature}
\newacro{DNN}{deep neural network}
\newacro{F0}{fundamental frequency}
\newacro{EER}{equal error rate}
\newacro{WER}{word error rate}
\newacro{PI}{personal information}
\newacro{PESQ}{perceptual evaluation of speech quality}
\newacro{MCD}{mel-cepstral distortion}
\newacro{MSE}{mean-squared error}
\newacro{NSF}{neural source-filter}
\newacro{AM}{acoustic model}
\newacro{VPC}{VoicePrivacy Challenge}
\newacro{ASV}{automated speaker verification}
\newacro{ASR}{automated speech recognition}
\newacro{TDNN}{time delay neural network}
\newacro{FPE}{Fine Pitch Error}
\newacro{GPE}{Gross pitch error}
\newacro{SI-SNR}{scale-invariant signal-to-noise ratio}
\newacro{STOI}{short-term objective intelligibility}
\begin{abstract}
Speaker anonymization systems continue to improve their ability to obfuscate the original speaker characteristics in a speech signal, but often create processing artifacts and unnatural sounding voices as a tradeoff. Many of those systems stem from the \ac{VPC} Baseline B1, using a neural vocoder to synthesize speech from an F0, x-vectors and bottleneck features-based speech representation. Inspired by this, we investigate the reproduction capabilities of the aforementioned baseline, to assess how successful the shared methodology is in synthesizing human-like speech. 
We use four objective metrics to measure speech quality, waveform similarity, and F0 similarity. Our findings indicate that both the speech representation and the vocoder introduces artifacts, causing an unnatural perception. 
\begin{subjeval}
A MUSHRA-like listening test on 18 subjects corroborate our findings, motivating 
\end{subjeval}
further research on the analysis and synthesis components of the \ac{VPC} Baseline B1.
\end{abstract}

\acresetall

\section{Introduction}
Numerous developments in the speech signal processing domain have rendered the collection of speech data as well as its adversarial utilization simpler \cite{tomashenko_introducing_2020}. As a result, voice privacy is an emerging issue in today's world. Many technical applications either require by law, or would benefit from, a preliminary processing to mitigate the risks to user privacy. In this regard, a \ac{VPC} has been organized to promote the development of voice anonymization systems via the introduced baselines, evaluation metrics and attack models, which are widely adopted by the researchers in the field.

Depending on the downstream task, i.e., the purpose the acquired speech signals shall serve, the anonymization procedure may be expected to preserve the prosody and the naturalness. One such use case is a psychiatric support context where the patients want to stay anonymous \cite{ingo_siegert_anonymprevent_2021}. However, the results from the \ac{VPC} 2020 and 2022 point out that none of the published systems up to date can achieve subjective naturalness scores on par with recorded human speech \cite{tomashenko_vpc_results_2020, tomashenko_vpc_results_2022}. Furthermore, our previous work utilizing contrastive systems revealed that using original x-vectors during synthesis surprisingly yields worse utility and an increase in the privacy \cite{Gaznepoglu:EUSIPCO23}. Therefore, in this work, we evaluate the speech resynthesis capabilities of the \ac{VPC} Baseline B1, using metrics from other domains, to understand if the speech representation and synthesis block shared across systems of multiple contestants have any improvement potential.

\section{Related work}
\subsection{VPC Baseline B1 and its derivatives}
\begin{figure}[ht]
    \centering
    \resizebox{\linewidth}{!}{\begingroup
\renewcommand{\baselinestretch}{1} \endlinechar=-1 \begin{tikzpicture}[ipe stylesheet]
  \draw[shift={(104, 768)}, xscale=0.6875, yscale=0.4]
    (0, 0) rectangle (64, -160);
  \draw[shift={(220, 700)}, yscale=0.4, -{>[ipe arrow small]}]
    (0, 0)
     -- (0, 20);
  \draw[shift={(64, 736)}, xscale=0.25, -{>[ipe arrow small]}]
    (0, 0)
     -- (160, 0);
  \node[ipe node, anchor=south west]
     at (70.28, 737) {Input};
  \node[ipe node, anchor=north west]
     at (67.374, 735) {Speech};
  \draw[-{>[ipe arrow small]}]
    (248, 716)
     -- (260, 716)
     -- (260, 724)
     -- (272, 724);
  \draw[-{>[ipe arrow small]}]
    (148, 736)
     -- (272, 736);
  \draw[-{>[ipe arrow small]}]
    (148, 760)
     -- (260, 760)
     -- (260, 748)
     -- (272, 748);
  \node[ipe node, anchor=center, font=\scriptsize]
     at (160, 740) {BN};
  \node[ipe node, anchor=center, font=\scriptsize]
     at (160, 764) {F0};
  \node[ipe node, anchor=west, font=\scriptsize]
     at (260, 708) {anonymized};
  \node[ipe node, anchor=center]
     at (126, 736) {ASR};
  \draw[shift={(108, 744.001)}, xscale=0.75, yscale=0.6667]
    (0, 0) rectangle (48, -24);
  \draw[shift={(108, 724)}, xscale=0.75, yscale=0.6667]
    (0, 0) rectangle (48, -24);
  \node[ipe node, anchor=center]
     at (126, 716) {ASV};
  \begin{scope}[shift={(0, 4)}]
    \draw[-{>[ipe arrow small]}]
      (352, 732)
       -- (392, 732);
    \node[ipe node, anchor=south west]
       at (353.97, 733) {Output};
    \node[ipe node, anchor=north west]
       at (355.077, 731) {Speech};
  \end{scope}
  \draw[shift={(272, 756.002)}, scale=1.6667]
    (0, 0) rectangle (48, -24);
  \node[ipe node, anchor=center]
     at (312, 748) {Neural Vocoder};
  \node[ipe node, anchor=center, font=\small]
     at (312, 736) {(AM-NSF /};
  \node[ipe node, anchor=center, font=\small]
     at (312, 724) {HiFi GAN etc.)};
  \node[ipe node, anchor=west, font=\scriptsize]
     at (260, 700) {x-vector};
  \node[ipe node, anchor=center, font=\small]
     at (126, 696) {Feature};
  \node[ipe node, anchor=center, font=\small]
     at (126, 688) {Extractors};
  \draw[shift={(191.999, 724)}, xscale=1.1667, yscale=0.6667]
    (0, 0) rectangle (48, -24);
  \node[ipe node, anchor=center]
     at (220, 716) {Anonymize};
  \begin{scope}[shift={(0, -4)}]
    \draw[shift={(108, 768)}, xscale=0.75, yscale=0.6667]
      (0, 0) rectangle (48, -24);
    \node[ipe node, anchor=center]
       at (126, 760) {F0 Ext.};
  \end{scope}
  \begin{scope}[shift={(0, 8)}]
    \node[ipe node, anchor=center, font=\small]
       at (224, 684) {X-vector pool};
    \node[ipe node, anchor=center, font=\scriptsize]
       at (192, 684) {\faDatabase};
    \draw[shift={(184, 692)}, yscale=0.6667]
      (0, 0) rectangle (72, -24);
  \end{scope}
  \begin{scope}[shift={(0, 4)}]
    \draw[shift={(148, 712)}, xscale=0.9167, -{>[ipe arrow small]}]
      (0, 0)
       -- (48, 0);
    \node[ipe node, anchor=center, font=\scriptsize]
       at (168, 716) {X-vector};
  \end{scope}
\end{tikzpicture}\endgroup}\\
    \caption{The {VPC} 2022 Baseline B1.a/b.}
    \label{fig:baseline}
    \vspace{-1.5em}
\end{figure}
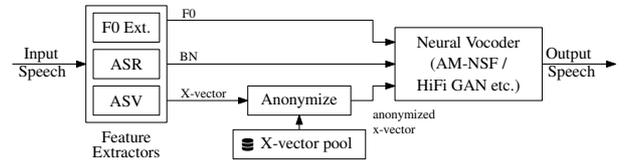

The VoicePrivacy Challenge Baseline B1 has been a source of inspiration to many challenge participants \cite{tomashenko_vpc_results_2020, tomashenko_vpc_results_2022}. The system \cite{fang_speaker_2019}, which consists of three feature extractors, an anonymization block, and a neural vocoder, is depicted in Figure \ref{fig:baseline}. The feature extractors and their purposes are outlined in the Table \ref{tab:features}.

\begin{table}[bh]
    \centering
    \caption{Extracted features per utterance. The quantity in parentheses indicates the resulting tensor shape. N: number of frames of an utterance. W: window size (ms), H: hop size (ms)}
    \vspace{-0.5em}
    \begin{tabular}{lll}
        \toprule
        \textbf{Feature (purpose)} & \textbf{Extractor} & \textbf{Properties} \\
        \cmidrule{1-1} \cmidrule{2-2} \cmidrule{3-3}
        \acs{F0} (Prosody) & YAAPT & (Nx1), W: \SI{35}{}, H: \SI{10}{} \\
        \acs{BN} (Verbal content) & TDNN-F & (Nx256), H: \SI{10}{} \\
         X-vector (Identity) & TDNN & (1x512)\\
        \bottomrule
    \end{tabular}
    \vspace{-0.5em}
    \label{tab:features}
\end{table}

More than 10 systems are proposed to improve the various aspects of the baseline over the last three years. Majority of these contributions target the anonymization block and keep the speech representation or the vocoder intact. Some however, such as \cite{champion_speaker_2020}, propose alternatives to the bottleneck features. For the speaker embedding, \cite{khamsehashari_voice_2022} proposes switching to ECAPA and \cite{meyer_speaker_2022} reports increased speaker representation capabilities when both ECAPA and x-vectors are used together. 

The \ac{NSF}, i.e., the neural vocoder, has also received some attention. The 2022 edition of the challenge included two vocoders (NSF-HiFiGAN and HiFiGAN) that directly predict the waveforms from the speech representation, discarding the \ac{AM} that was present in the 2020 baseline. Works such as \cite{meyer_anonymizing_2022, meyer_speaker_2022} use the IMS Toucan toolkit that provides a modular neural vocoder and utilize local energy in addition to F0 for further prosody control.

\subsection{Evaluation of voice conversion systems}

The voice anonymization problem, especially the way \ac{VPC} framework treats it, has some similarities to the voice conversion problem. An overview on voice conversion mentions intrusive metrics like \ac{PESQ} and \ac{MCD} to evaluate synthesized speech quality \cite{sisman_overview_2020}. In our study, the availability of the reference signals lets us employ such methods. 
\begin{torchsquim}
Recently, torchaudio-SQUIM was proposed to estimate metrics such as \ac{PESQ} on synthesized speech without needing a reference \cite{kumar_torchaudio-squim_2023}.
\end{torchsquim}

\subsection{Evaluation of voice anonymization systems}

The \ac{VPC} framework introduced objective and subjective metrics to evaluate different aspects of the anonymized speech signals \cite{tomashenko_vpc_evalplan_2022}. The \ac{WER}, whose lower values indicate a better utility, is measured by an \ac{ASR} system. An \ac{ASV} system is used to measure the anonymization success, where higher \ac{EER} values indicate better anonymization. Prosody retention to a certain extent is ensured by a lower bound on F0 correlation and finally, a gain of voice distinctiveness measures whether the speaker diversity of the input speech datasets are preserved by the anonymization process. However, none of the introduced objective metrics can successfully measure the perceived naturalness thus the challenge organizers have resorted to a subjective evaluation of the utterances \cite{tomashenko_vpc_evalplan_2022}.

The \ac{VPC} community uses contrastive systems \cite{mawalim_x-vector_2020, champion_speaker_2020, Gaznepoglu:EUSIPCO23}, an idea similar to the ablation studies performed by the machine learning community. A contrastive system is a marginally different configuration of the anonymization system that provides further insights into how different modules thereof contribute to the performance. The cited studies use \ac{VPC} metrics to assess the privacy versus utility (\ac{ASR} scenario) tradeoff and also reported that synthesis with original set of features cause an increase to \ac{EER} as well as to \ac{WER}, hinting that some artifacts are introduced by the analysis-synthesis pipeline. 

To conclude, the existing objective metrics of the \ac{VPC} do not account for naturalness. The alternative, subjective listening tests, are non-ideal because they are time-consuming and costly. Furthermore, the evaluation methods in the VoicePrivacy literature are not capable of detecting abnormal behaviors in time, which in our opinion is necessary to find what causes unnatural outputs. To go beyond the contrastive system studies with \ac{EER} and \ac{WER}, we decided to investigate whether intrusive metrics
\begin{torchsquim}
and their non-intrusive estimates
\end{torchsquim}
 could be exploited.
 
\section{Methodology}

\subsection{Dataset}

\begin{table}[b]
    \centering
    \vspace{-0.5em}
    \caption{\ac{VPC} data subsets \cite{tomashenko_vpc_evalplan_2022} utilized in this work. \#F, \#M: number of unique female/male speakers}
    \vspace{-5pt}
    \begin{tabular}{lrrrr}
        \toprule
        \textbf{Subset Name} & \makecell[c]{\textbf{\#F}} & \makecell[c]{\textbf{\#M}} & \makecell[c]{\textbf{\#Utterances}}\\
        \midrule
        libri-test-\{enrolls,trials\} & 15  & 15    & 1934  \\
        vctk-test-\{enrolls,trials\}  & 15  & 15    & 12048 \\
        \bottomrule
    \end{tabular}
    \label{tab:dataset}
\end{table}

We use the \ac{VPC} datasets \texttt{libri}-* and \texttt{vctk}-* for our evaluations. A summary of their content is provided in Tab. \ref{tab:dataset}. These datasets are resynthesized using the systems in Table \ref{tab:systems}. The system \texttt{B1b-spk} is the same as the 2022 baseline, except it uses the original speaker-level x-vectors for synthesis. The system \texttt{B1b-utt}, using the utterance-level x-vectors for synthesis, imitates the training conditions of the neural vocoders. Both systems were trained using HiFiGAN discriminators. The system \texttt{joint-hifigan-spk} denotes the alternative vocoder (HiFiGAN \cite{kong_hifi-gan_2020}) provided by the \ac{VPC} organizers. The system \texttt{am-nsf-spk} is the baseline used in 2020, that features an additional autoregressive \ac{AM} that converts the speech representation into mel-spectrograms. The system \texttt{mel-nsf-spk} bypasses the \ac{AM} and performs synthesis using the mel-spectrograms computed from original utterances, also referred to as \textit{copy-synthesis} \cite{wang_spoofed_2023}. We feature both the PyTorch variant (denoted with a suffix \texttt{-pt}) and the C-based implementation utilized in \ac{VPC} 2020. We also included an anchor equivalent \texttt{mel-nsf-spk-4k} that sets the mel-spectogram values for frequency bands with $f_c > $ \si{4}{kHz} to zero.

A number of pre-processing steps are performed before the evaluation. Systems we evaluate introduce different amounts of delay, so we align the outputs with the references using cross-correlation. Many of the utterances contain silence, as well as some pauses, hence we ran Silero voice activity detection \cite{Silero_VAD} on the references and computed the metrics on the segments with voice. Also, a number of utterances were visually inspected to ensure that the synthesis procedure preserves the loudness, which could bias the evaluation scores \cite{qiao_case_2008}.

\begin{table}[t]
    \centering
    \caption{Systems evaluated in this paper. The x-vectors are not anonymized to assess the resynthesis capability. Vocoders are \ac{VPC} PyTorch implementations, unless noted in the table.}
    \vspace{-0.5em}
    \begin{tabular}{lcc}
    \toprule
    \textbf{ID} & \textbf{X-vector} & \textbf{Vocoder}\\
    \midrule
    mel-nsf-pt-spk    & speaker-level & \ac{NSF} \\
    mel-nsf-spk       & speaker-level & (C-based) \ac{NSF} \\
    mel-nsf-spk-4k    & speaker-level & (C-based) \ac{NSF} \\
    am-nsf-spk        & speaker-level & (C-based) \ac{AM} + \ac{NSF} \\
    B1b-utt           & utterance-level & joint \ac{NSF} (+HiFiGAN-D)\\
    B1b-spk           & speaker-level & joint \ac{NSF} (+HiFiGAN-D)\\
    joint-hifigan-spk & speaker-level & joint HiFiGAN\\
    \bottomrule
\end{tabular}
    \label{tab:systems}
    \vspace{-1.25em}
\end{table}

\subsection{Objective evaluation metrics}\label{ssec:obj_eval}
We adopt four different objective metrics to evaluate the resynthesis capabilities. These metrics are all intrusive, meaning that their computation requires access to a reference signal.

\subsubsection{\texorpdfstring{\Acf{MCD}}{}}
\ac{MCD} is used to measure the signal similarity in a perceptual sense. The implementation we use is provided by \cite{sternkopf_mel-cepstral-distance_2022}.
\subsubsection{\texorpdfstring{\Acf{SI-SNR}}{}}
\ac{SI-SNR} is used to measure the signal similarity \cite{roux_sdr_2019}. The reference signal is first projected on the estimated signal, to obtain a scaling coefficient. Then the signal-to-noise ratio is computed. The implementation we use is a NumPy port of \cite{detlefsen_torchmetrics_2022}.
\subsubsection{\texorpdfstring{\Acf{PESQ}}{}}
\ac{PESQ} is an intrusive measure introduced by ITU to predict the subjective speech quality evaluations. We use the implementation in \texttt{python-pesq} \cite{wang_pesq_2022}.
\subsubsection{\texorpdfstring{\Acf{GPE}}{}}
\ac{GPE} is a metric for F0 extractor evaluation. In our work, we use it to compare the synthesized F0 to the original. We adopt the definition in \eqref{eq:gpe}, also used in a previous work of us \cite{Gaznepoglu:EUSIPCO23}.
\begin{equation}\label{eq:gpe}
    \text{GPE: } \frac{\text{num. of frames whose error}>20\%}{\text{num. of correctly identified voiced frames}}
\end{equation}

\ac{MCD}, \ac{SI-SNR} and \ac{GPE} have the advantage that they could be computed on smaller segments.

\begin{torchsquim}
\subsection{torchaudio-SQUIM}\label{ssec:squim}

In addition to the intrusive metrics, we also tested the torchaudio-SQUIM \cite{kumar_torchaudio-squim_2023}, which provides non-intrusive estimates of the intrusive metrics. We use their \ac{PESQ} prediction and report numbers for all the classes as well as for the reference signals. If these estimates correlate well with their intrusive complements or with user preferences, SQUIM could be also tested for evaluating anonymized speech.

\end{torchsquim}



\begin{subjeval}
\subsection{Subjective listening test}
We conducted a MUSHRA-like listening test on 18 subjects of varying listening test experiences, using webMUSHRA software \cite{schoeffler_webmushra_2018}. We randomly picked eight utterances from \texttt{libri-test} and six from \texttt{vctk-test}, (7 male and 7 female speakers, utterance lengths between $[5.5, 8]$ seconds), which are available at \footnote{\url{https://audiolabs-erlangen.de/resources/2023-VPC-resynth-eval}}. The users are presented each synthesis output and asked to rate the naturalness using the following prompt, inspired by the \ac{VPC} subjective test \cite{tomashenko_vpc_evalplan_2022}.\\

\begin{quoting}
\noindent You will listen to a series of audio samples, comprising of both original recordings (referred to as reference) and versions that have been resynthesized using different neural vocoders, resulting in varying degrees of artifacts. Your task is to rate the naturalness of each recording.

\noindent\textbf{Naturalness:} Please judge how much audio degradation you can hear in each file. You need to select a score in the interval $[0, 100]$, where higher numbers correspond to a more natural sounding audio, a $0$ corresponding to 'severely degraded' and a $100$ to 'no degradation at all'. For this score please only consider the sound characteristics and not the content. Also note that the reference contains some background noise. Finally, the deviations from original speaker's voice also count as degradations. 
\end{quoting}
\end{subjeval}

\section{Results and Discussion}
\subsection{Objective evaluation}

\begin{figure}[ht]
    \centering
    \begingroup
\renewcommand{\baselinestretch}{1} \endlinechar=-1 \input{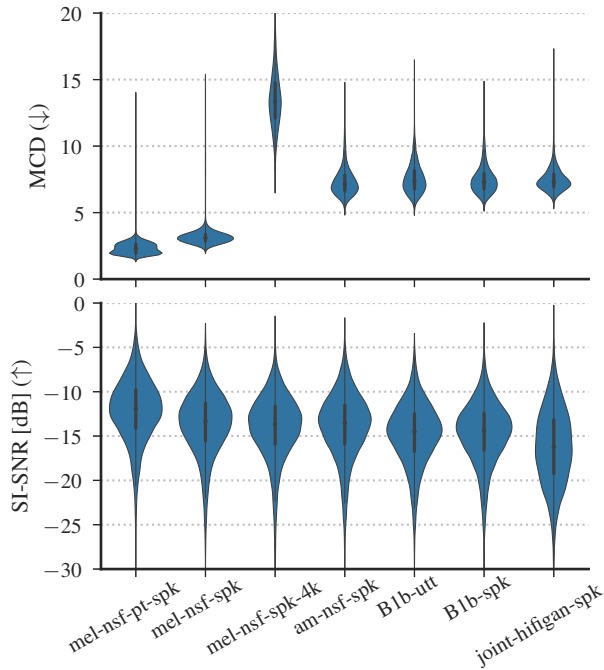}\endgroup
    \vspace{-2em}
    \caption{Evaluation results for signal similarity metrics.}
    \label{fig:obj_eval_sig}
    \vspace{-1em}
\end{figure}

\subsubsection{Signal similarity metrics}

Figure \ref{fig:obj_eval_sig} depicts the \ac{SI-SNR} and \ac{MCD} results. We saw no significant differences during visual inspection of the "per-dataset" and "per-gender" distributions. Therefore we display averages over datasets and gender instead.

Copy synthesis, e.g., \texttt{mel-nsf-spk}, outperformed the others, but \texttt{mel-nsf-pt-spk} and \texttt{mel-nsf-spk}, two implementations of the same system, behaved differently. PyTorch copy-synthesis achieved better \ac{SI-SNR} and \ac{MCD}. The anchor, i.e., \texttt{mel-nsf-spk-4k}, attained comparable \ac{SI-SNR} but the worst \ac{MCD}. Other vocoders attained a similar \ac{MCD}, standing between the copy synthesis and the anchor. We interpret the discrepancy between \texttt{mel-nsf} and synthesis from the representation as a sign of inadequacies of the utilized speech representation, resulting in some further information loss on top of the artifacts due to \ac{NSF}. \texttt{am-nsf-spk} performed slightly better than other vocoders, indicating the \ac{AM} contributed to the resynthesis performance. 

\subsubsection{F0 similarity}

\begin{figure}[hbt]
    \centering
    \begingroup
\renewcommand{\baselinestretch}{1} \endlinechar=-1 \input{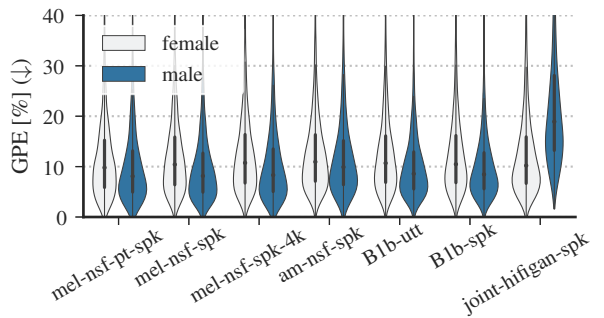}\endgroup
    \vspace{-2em}
    \caption{F0 similarity evaluation.}
    \label{fig:obj_eval_gpe}
    \vspace{-1em}
\end{figure}

In a similar manner, Figure \ref{fig:obj_eval_gpe} depicts the \ac{GPE} results. Behavior across female and male recordings are shown this time.

\ac{NSF}-based systems maintained a certain standard in terms of F0 preservation, due to the source-filter model. HiFiGAN takes some extra liberty whilst synthesizing the signals, thus attained significantly higher \ac{GPE} and this probably explains why it attained the worst \ac{SI-SNR} too. An interesting outcome is that female speech has slightly higher \ac{GPE} for \ac{NSF}, but HiFiGAN corrupts pitch significantly more for male speakers. 

\subsubsection{PESQ and torchaudio-SQUIM}

\begin{figure*}[htb]
    \centering
    \begingroup
\renewcommand{\baselinestretch}{1} \endlinechar=-1 \input{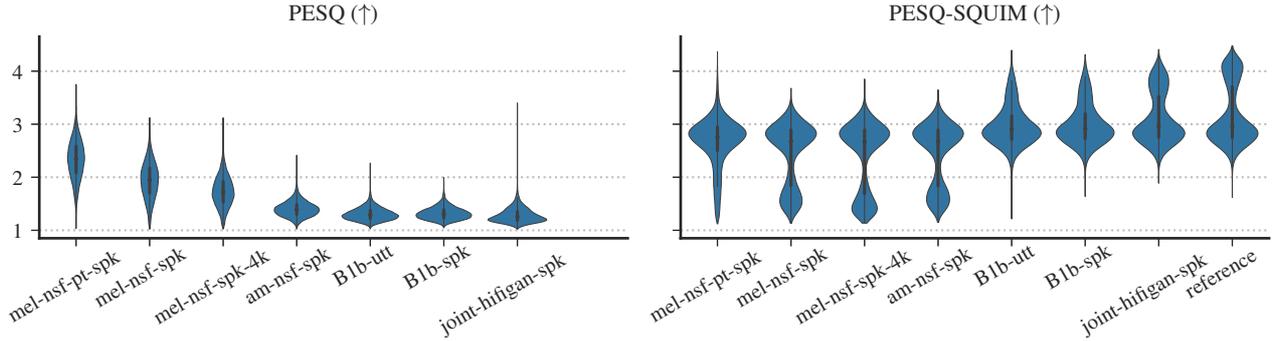}\endgroup
    \vspace{-1em}
    \caption{Evaluation results for PESQ and torchaudio-SQUIM estimate of PESQ.}
    \vspace{-1em}
    \label{fig:obj_eval_pesq}
\end{figure*}

Finally, we compare the PESQ computations as well as PESQ estimates by torchaudio-SQUIM in Figure \ref{fig:obj_eval_pesq}.

\ac{PESQ} with respect to the reference (left) shows a similar, but a more pronounced version of the trend in the \ac{SI-SNR} plots. PyTorch copy-synthesis, i.e., \texttt{mel-nsf-pt-spk}, attained the best \ac{PESQ} scores. The anchor performed better than the variants that synthesize from the speech representation (e.g., \texttt{B1b}). \texttt{am-nsf-spk} performed slightly better than other vocoders, again hinting the joint AM-NSF approach introduced in 2022 causing minor degradation. \texttt{joint-hifigan-spk} performed the worst. On metrics that take the perceptual aspects into account, such as MCD and PESQ, \texttt{B1b-spk} performed better than \texttt{B1b-utt}, which imitates the vocoder training scenario. This may indicate an underfit. A number of factors could have caused this, such as an insufficient representation complexity, lack of augmentation (augmenting x-vectors might help the vocoder to better learn the neighboring relations of the x-vectors) or may simply indicate that the training procedure has been cut off too early. The \ac{NSF} was trained using L1 loss \cite{fang_speaker_2019} on the magnitude spectrogram, which could be substituted with a perceptual loss to improve the performance.

Interestingly, the PESQ scores exhibit a greater inter-utterance variance for \texttt{mel-nsf-spk} variants. Additional investigations are required to understand this phenomenon. In particular it is crucial to understand if a confounding variable affects the scores, as previously shown by \cite{qiao_case_2008} with PESQ for factors such as loudness and alignment.

\begin{torchsquim}
    Torchaudio-SQUIM estimates of PESQ showed a different behavior. Systems \texttt{joint-hifigan-spk}, \texttt{B1b-utt} and \texttt{B1b-spk} achieved better performance with torchaudio-SQUIM evaluation. The systems have the HiFiGAN discriminators in common, which possibly explains the outcome. Some systems, e.g., \texttt{am-nsf-spk}, showed an unexpected bimodal distribution that is not explained by the gender or the dataset, which needs further investigations. The SQUIM estimates for the reference signals, depicted by the right-most violin plot in the Figure \ref{fig:obj_eval_pesq}, again show a bimodal behavior. 

\end{torchsquim}

\begin{subjeval}
\subsection{Subjective listening test}

The listening test responses are filtered such that the answers for an utterance, whose reference was rated with less than $90$ points, are removed. This results in at least $14$ subjects rating each utterance. The ratings are presented in Figure \ref{fig:subj_eval}.

\begin{figure}[htb]
    \centering
    \begingroup
\renewcommand{\baselinestretch}{1} \endlinechar=-1 \input{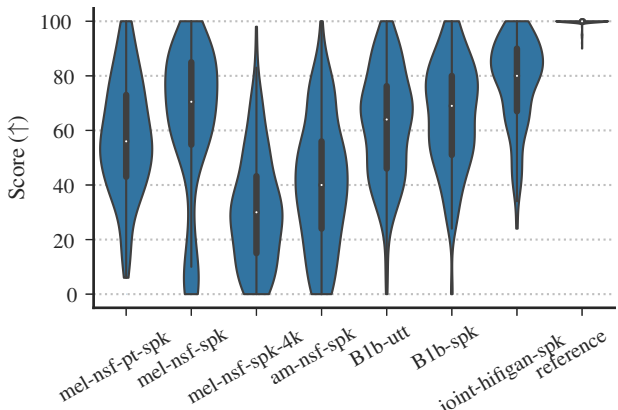}\endgroup\\
    \vspace{-1em}
    \caption{The subjective evaluation results.}
    \label{fig:subj_eval}
    \vspace{-1em}
\end{figure}

Subjects reported that some \texttt{mel-nsf-spk} utterances had a severe muffling effect, often at their beginnings, rendering the part of the utterance completely unintelligible. In contrast, the \texttt{mel-nsf-pt-spk} was reported to suffer from random impulsive artifacts, somewhat like a "sizzling frying pan constantly accompanying the recordings". \texttt{joint-hifigan-nsf} was reported to change the accents, "Americanizing" the voices, and the identity perception was different to what reference or other systems evoked. Otherwise the speech was reported to sound most human-like.

Now turning to the analysis of the gathered scores, most subjects were able to identify the reference stimuli and grade accordingly. Removed answers constitute less than 10\% of the acquired data. The anchor \texttt{mel-nsf-spk-4k} was rated the lowest whereas \texttt{joint-hifigan-spk} was rated the best, except it compromises on the speaker identity. Systems other than \texttt{joint-hifigan-spk} exhibited a higher inter-utterance variance. Notwithstanding the few utterances with unintelligible segments causing a second modality at the bottom, the copy-synthesis, \texttt{mel-nsf-spk} was rated the second best, followed by \texttt{B1b} variants. \texttt{mel-nsf-pt-spk} and \texttt{am-nsf-spk} were rated only slightly better than the anchor. 

\subsubsection{Predictability of the subjective test scores}

Intrusive metrics could not predict the outcome that \texttt{joint-hifigan-nsf} would be perceived the most natural, \texttt{B1b} performing better than \texttt{am-nsf-spk} and \texttt{mel-nsf-spk} outperforming \texttt{mel-nsf-pt-spk}. Only evaluation we ran that anticipated this outcome was torchaudio-SQUIM. We conclude that, the reference being available causes the intrusive evaluation to focus on the differences in signals that our subjects did not consider. Among the objective metrics, MCD was relatively successful.

Comparison of the PyTorch-based \texttt{mel-nsf-pt-spk} and C-based \texttt{mel-nsf-spk}, our subjects rated the latter better. The subjects penalized non-stationary artifacts less. To conclude, even though the objective metrics we utilize in this paper contribute to understanding how the blocks interact, these are not sufficient to explain the subject preferences completely.
\end{subjeval}

\subsection{Future work}
We think it would be worthwhile to study the effects of using additional speaker embeddings such as ECAPA \cite{desplanques_ecapa-tdnn_2020}, as multiple systems in the literature utilized it and performed well in the \ac{VPC} 2022. Part of ECAPA's success comes from a better temporal pooling strategy using attention. However, \ac{VPC} simply uses temporal averaging to obtain the utterance-level x-vectors, and mere utterance averages to obtain speaker-level x-vectors, so modifications to these aspects are worth investigating.

Some of the metrics we used, e.g., MCD, GPE and SI-SNR allow computation on very small segments, unlike PESQ. The time segments with the reported muffling effect could be automatically located with these and further analysis could be conducted. Also for these metrics, different temporal pooling strategies could be experimented with.

\section{Conclusion}

In this paper, we have investigated the reproduction capabilities of the VoicePrivacy Challenge Baseline B1 by utilizing a diverse set of objective evaluation metrics. Our subjective and objective evaluation results indicate that the copy synthesis scores better than the synthesis from representations, likely indicating the speech representation is causing additional information loss and yielding unnatural sounding output. Previous studies found that a more recent speaker embedding could help improve the anonymization performance, and our results hint that it could also improve the synthesized speech quality. In addition, the vocoder training scheme may benefit from a number of changes to bolster its understanding of the speaker embedding space.

The objective metrics we utilize in this work show limited effectiveness in evaluating the system behavior for anonymization, primarily because they are intrusive and no references are available for anonymization, and metrics we evaluated partially align with the listening test subject preferences. Torchaudio-SQUIM's PESQ implementation performed relatively well, and it does not require a reference, so the voice anonymization evaluations may benefit from it. 

\printbibliography[heading=bibnumbered]

@inproceedings{Gaznepoglu:EUSIPCO23,	
	author = {Ünal Ege Gaznepoğlu and Nils Peters},
	booktitle = {the forthcoming Proc. of the 31st European Signal Processing Conference (EUSIPCO)},
	title = {Deep Learning-based F0 Synthesis for Speaker Anonymization},
	address = {Helsinki, Finland},
	year = {2023}}

@misc{Silero_VAD,
  author = {Alexander Veysov},
  title = {Silero {VAD}: pre-trained enterprise-grade Voice Activity Detector {(VAD)}},
  year = {2021},
  publisher = {GitHub},
  journal = {GitHub repository},
  howpublished = {\url{https://github.com/snakers4/silero-vad}},
  commit = {insert_some_commit_here},
  email = {hello@silero.ai}
}

@inproceedings{meyer_anonymizing_2022,
	title = {Anonymizing Speech with Generative Adversarial Networks to Preserve Speaker Privacy},
	url = {http://arxiv.org/abs/2210.07002},
	booktitle = {Proc. {IEEE} Spoken Lang. Tech. Workshop ({SLT})},
	publisher = {{arXiv}},
	author = {Meyer, Sarina and Tilli, Pascal and Denisov, Pavel and Lux, Florian and Koch, Julia and Vu, Ngoc Thang},
	urldate = {2022-12-05},
	date = {2022-10-20},
	eprinttype = {arxiv},
	eprint = {2210.07002 [cs, eess]},
	note = {00000 },
}

@inproceedings{qiao_case_2008,
	location = {Cannes, France},
	title = {Case study of {PESQ} performance in live wireless mobile {VoIP} environment},
	isbn = {978-1-4244-2643-0},
	url = {http://ieeexplore.ieee.org/document/4699880/},
	doi = {10.1109/PIMRC.2008.4699880},
	eventtitle = {2008 {IEEE} 19th International Symposium on Personal, Indoor and Mobile Radio Communications ({PIMRC})},
	pages = {1--6},
	booktitle = {{IEEE} 19th Intl. Symp. on Personal, Indoor and Mobile Radio Communications},
	publisher = {{IEEE}},
	author = {Qiao, Zizhi and Sun, Lingfen and Ifeachor, Emmanuel},
	urldate = {2023-06-14},
	date = {2008-09},
	langid = {english},
	note = {00028},
}

@inproceedings{khamsehashari_voice_2022,
	title = {Voice Privacy - leveraging multi-scale blocks with {ECAPA}-{TDNN} {SE}-Res2NeXt extension for speaker anonymization},
	url = {https://www.isca-speech.org/archive/spsc_2022/khamsehashari22_spsc.html},
	doi = {10.21437/SPSC.2022-8},
	eventtitle = {2nd Symposium on Security and Privacy in Speech Communication},
	pages = {43--48},
	booktitle = {2nd Symp. on Security and Privacy in Speech Communication},
	publisher = {{ISCA}},
	author = {Khamsehashari, Razieh and Sinha, Yamini and Hintz, Jan and Ghosh, Suhita and Polzehl, Tim and Franzreb, Clarlos and Stober, Sebastian and Siegert, Ingo},
	urldate = {2023-06-06},
	date = {2022-09-23},
	langid = {english},
	note = {00000},
}

@inproceedings{kong_hifi-gan_2020,
	title = {{HiFi}-{GAN}: Generative Adversarial Networks for Efficient and High Fidelity Speech Synthesis},
	url = {http://arxiv.org/abs/2010.05646},
	doi = {10.48550/arXiv.2010.05646},
	shorttitle = {{HiFi}-{GAN}},
	eventtitle = {Proc. {NeurIPS} Conf.},
	publisher = {{arXiv}},
	author = {Kong, Jungil and Kim, Jaehyeon and Bae, Jaekyoung},
	urldate = {2023-08-01},
	date = {2020-10-23},
}

@inproceedings{mawalim_x-vector_2020,
	title = {X-Vector Singular Value Modification and Statistical-Based Decomposition with Ensemble Regression Modeling for Speaker Anonymization System},
	url = {https://www.isca-speech.org/archive/interspeech_2020/mawalim20_interspeech.html},
	doi = {10.21437/Interspeech.2020-1887},
	eventtitle = {Interspeech 2020},
	pages = {1703--1707},
	booktitle = {Proc. Interspeech Conf.},
	publisher = {{ISCA}},
	author = {Mawalim, Candy Olivia and Galajit, Kasorn and Karnjana, Jessada and Unoki, Masashi},
	urldate = {2023-07-28},
	date = {2020-10-25},
	langid = {english},
	note = {00019},
}

@online{wang_pesq_2022,
	title = {pesq: Python Wrapper for {PESQ} Score (narrow band and wide band)},
	rights = {{MIT} License},
	url = {https://github.com/ludlows/python-pesq},
	shorttitle = {pesq},
	author = {Wang, Miao and Boedekker, Christoph and Dantas, Rafael G. and Seelan, Ananda},
	urldate = {2023-06-13},
	date = {2022},
	note = {00000},
}

@inproceedings{desplanques_ecapa-tdnn_2020,
	title = {{ECAPA}-{TDNN}: Emphasized Channel Attention, Propagation and Aggregation in {TDNN} Based Speaker Verification},
	url = {http://arxiv.org/abs/2005.07143},
	doi = {10.21437/Interspeech.2020-2650},
	shorttitle = {{ECAPA}-{TDNN}},
	pages = {3830--3834},
	booktitle = {Proc. Interspeech Conf.},
	author = {Desplanques, Brecht and Thienpondt, Jenthe and Demuynck, Kris},
	urldate = {2023-06-13},
	date = {2020-10-25},
	eprinttype = {arxiv},
	eprint = {2005.07143 [cs, eess]},
	note = {00629 },
}

@inproceedings{meyer_speaker_2022,
	title = {Speaker Anonymization with Phonetic Intermediate Representations},
	url = {http://arxiv.org/abs/2207.04834},
	doi = {10.48550/arXiv.2207.04834},
	booktitle = {Proc. Interspeech Conf.},
	publisher = {{arXiv}},
	author = {Meyer, Sarina and Lux, Florian and Denisov, Pavel and Koch, Julia and Tilli, Pascal and Vu, Ngoc Thang},
	urldate = {2022-12-01},
	date = {2022-07-11},
	eprinttype = {arxiv},
	eprint = {2207.04834 [cs, eess]},
	note = {00005 },
}

@online{sternkopf_mel-cepstral-distance_2022,
	title = {mel-cepstral-distance},
	rights = {{MIT}},
	url = {https://github.com/jasminsternkopf/mel_cepstral_distance},
	author = {Sternkopf, Jasmin and Taubert, Stefan},
	urldate = {2023-06-13},
	date = {2022},
	doi = {10.5281/zenodo.7044405},
	note = {00000 },
}

@online{tomashenko_vpc_results_2022,
	title = {The {VoicePrivacy} 2022 Challenge Results},
	url = {https://www.voiceprivacychallenge.org/results-2022/docs/VoicePrivacy_2022_Challenge___Natalia_Tomashenko.pdf},
	author = {Tomashenko, Natalia and Wang, Xin and Miao, Xiaoxiao and Nourtel, Hubert and Champion, Pierre and Todisco, Massimiliano and Vincent, Emmanuel and Evans, Nicholas and Yamagishi, Junichi and Bonastre, Jean-Francois and Panariello, Michele},
	date = {2022},
	langid = {english},
	note = {00000},
}

@inproceedings{kumar_torchaudio-squim_2023,
	title = {Torchaudio-Squim: Reference-Less Speech Quality and Intelligibility Measures in Torchaudio},
	doi = {10.1109/ICASSP49357.2023.10096680},
	shorttitle = {Torchaudio-Squim},
	eventtitle = {{ICASSP} 2023 - 2023 {IEEE} International Conference on Acoustics, Speech and Signal Processing ({ICASSP})},
	pages = {1--5},
	booktitle = {Proc. {IEEE} Intl. Conf. on Acoustics, Speech and Signal Processing ({ICASSP})},
	author = {Kumar, Anurag and Tan, Ke and Ni, Zhaoheng and Manocha, Pranay and Zhang, Xiaohui and Henderson, Ethan and Xu, Buye},
	date = {2023-06},
	note = {00000},
}

@inproceedings{roux_sdr_2019,
	title = {{SDR} - half-baked or well done?},
	url = {http://arxiv.org/abs/1811.02508},
	booktitle = {Proc. {IEEE} Intl. Conf. on Acoustics, Speech and Signal Processing ({ICASSP})},
	publisher = {{arXiv}},
	author = {Roux, Jonathan Le and Wisdom, Scott and Erdogan, Hakan and Hershey, John R.},
	urldate = {2023-05-18},
	date = {2019},
	langid = {english},
	eprinttype = {arxiv},
	eprint = {1811.02508 [cs, eess]},
	note = {00745 },
}

@inproceedings{wang_spoofed_2023,
	title = {Spoofed training data for speech spoofing countermeasure can be efficiently created using neural vocoders},
	url = {http://arxiv.org/abs/2210.10570},
	doi = {10.48550/arXiv.2210.10570},
	booktitle = {Proc. {IEEE} Intl. Conf. on Acoustics, Speech and Signal Processing ({ICASSP})},
	publisher = {{arXiv}},
	author = {Wang, Xin and Yamagishi, Junichi},
	urldate = {2023-06-08},
	date = {2023-02-22},
	eprinttype = {arxiv},
	eprint = {2210.10570 [cs, eess]},
	note = {00002 },
}

@online{tomashenko_vpc_evalplan_2022,
	title = {2nd {VoicePrivacy} Challenge Evaluation Plan},
	url = {https://arxiv.org/abs/2203.12468},
	author = {Tomashenko, Natalia and Wang, Xin and Miao, Xiaoxiao and Nourtel, Hubert and Todisco, Massimiliano and Vincent, Emmanuel and Evans, Nicholas and Bonastre, Jean-François},
	date = {2022},
	langid = {english},
	note = {00031
Type: online},
}

@inproceedings{tomashenko_introducing_2020,
	location = {Shanghai, China},
	title = {Introducing the {VoicePrivacy} Initiative},
	url = {http://arxiv.org/abs/2005.01387},
	eventtitle = {Interspeech},
	booktitle = {Proc. Interspeech Conf.},
	author = {Tomashenko, Natalia and Srivastava, Brij Mohan Lal and Wang, Xin and Vincent, Emmanuel and Nautsch, Andreas and Yamagishi, Junichi and Evans, Nicholas and Patino, Jose and Bonastre, Jean-François and Noé, Paul-Gauthier and Todisco, Massimiliano},
	urldate = {2021-02-18},
	date = {2020-10-25},
	langid = {english},
	eprinttype = {arxiv},
	eprint = {2005.01387},
	note = {00083 },
}

@article{detlefsen_torchmetrics_2022,
	title = {{TorchMetrics} - Measuring Reproducibility in {PyTorch}},
	volume = {7},
	issn = {2475-9066},
	url = {https://joss.theoj.org/papers/10.21105/joss.04101},
	doi = {10.21105/joss.04101},
	pages = {4101},
	number = {70},
	journaltitle = {Journal of Open Source Software},
	author = {Detlefsen, Nicki Skafte and Borovec, Jiri and Schock, Justus and Jha, Ananya Harsh and Koker, Teddy and Liello, Luca Di and Stancl, Daniel and Quan, Changsheng and Grechkin, Maxim and Falcon, William},
	urldate = {2023-06-13},
	date = {2022-02-11},
	langid = {english},
	note = {00012},
}

@online{ingo_siegert_anonymprevent_2021,
	title = {{AnonymPrevent} - {AI}-based Improvement of Anonymity for Remote Assessment, Treatment and Prevention against Child Sexual Abuse},
	url = {https://forschung-sachsen-anhalt.de/project/anonymprevent-based-improvement-anonymity-25374},
	author = {{Ingo Siegert} and {Sebastian Stober}},
	urldate = {2023-06-13},
	date = {2021},
	note = {00000},
}

@inproceedings{fang_speaker_2019,
	title = {Speaker anonymization using x-vector and neural waveform models},
	url = {http://arxiv.org/abs/1905.13561},
	booktitle = {Proc. 10th {ISCA} Speech Synthesis Workshop},
	author = {Fang, Fuming and Wang, Xin and Yamagishi, Junichi and Echizen, Isao and Todisco, Massimiliano and Evans, Nicholas and Bonastre, Jean-Francois},
	urldate = {2021-02-18},
	date = {2019-05-29},
	eprinttype = {arxiv},
	eprint = {1905.13561},
	note = {00061 },
}

@article{schoeffler_webmushra_2018,
	title = {{webMUSHRA} — A Comprehensive Framework for Web-based Listening Tests},
	volume = {6},
	issn = {2049-9647},
	url = {https://openresearchsoftware.metajnl.com/article/10.5334/jors.187/},
	doi = {10.5334/jors.187},
	pages = {8},
	number = {1},
	journaltitle = {Journal of Open Research Software},
	shortjournal = {{JORS}},
	author = {Schoeffler, Michael and Bartoschek, Sarah and Stöter, Fabian-Robert and Roess, Marlene and Westphal, Susanne and Edler, Bernd and Herre, Jürgen},
	urldate = {2023-05-31},
	date = {2018-02-05},
	langid = {english},
	note = {00154},
}

@article{tomashenko_vpc_results_2020,
	title = {The {VoicePrivacy} 2020 Challenge: Results and findings},
	volume = {74},
	issn = {08852308},
	url = {http://arxiv.org/abs/2109.00648},
	doi = {10.1016/j.csl.2022.101362},
	shorttitle = {The {VoicePrivacy} 2020 Challenge},
	pages = {101362},
	journaltitle = {Computer Speech \& Language},
	shortjournal = {Computer Speech \& Language},
	author = {Tomashenko, Natalia and Wang, Xin and Vincent, Emmanuel and Patino, Jose and Srivastava, Brij Mohan Lal and Noé, Paul-Gauthier and Nautsch, Andreas and Evans, Nicholas and Yamagishi, Junichi and O'Brien, Benjamin and Chanclu, Anaïs and Bonastre, Jean-François and Todisco, Massimiliano and Maouche, Mohamed},
	urldate = {2022-06-15},
	date = {2022-07},
	langid = {english},
	eprinttype = {arxiv},
	eprint = {2109.00648 [cs, eess]},
	note = {00015 },
}

@article{sisman_overview_2020,
	title = {An Overview of Voice Conversion and Its Challenges: From Statistical Modeling to Deep Learning},
	volume = {29},
	issn = {2329-9304},
	doi = {10.1109/TASLP.2020.3038524},
	shorttitle = {An Overview of Voice Conversion and Its Challenges},
	pages = {132--157},
	journaltitle = {{IEEE}/{ACM} Transactions on Audio, Speech, and Language Processing},
	author = {Sisman, B. and Yamagishi, J. and King, S. and Li, H.},
	date = {2020},
	note = {00000 
Conference Name: {IEEE}/{ACM} Transactions on Audio, Speech, and Language Processing},
}

@inproceedings{champion_speaker_2020,
	title = {Speaker information modification in {VoicePrivacy} 2020 toolchain},
	url = {https://hal.archives-ouvertes.fr/hal-02995855},
	booktitle = {{VoicePrivacy} Challenge Submission},
	author = {Champion, Pierre and Jouvet, Denis and Larcher, Anthony},
	urldate = {2021-06-22},
	date = {2020},
	note = {00013},
}

\end{document}